\documentclass[aps,amsmath,prb,twocolumn,a4paper,showpacs]{revtex4}
\usepackage{graphicx}

\newcommand{\beq}{\begin{equation}}
\newcommand{\eneq}{\end{equation}}
\newcommand{\bea}{\begin{eqnarray}} 
\newcommand{\enea}{\end{eqnarray}}
\newcommand{\bean}{\begin{eqnarray*}}
\newcommand{\eean}{\end{eqnarray*}}
\newcommand{\nn}{\nonumber}
\newcommand{\met}{\frac{1}{2}}  
\newcommand{\drond}[1]{\frac{\partial}{\partial #1} }
\newcommand{\EE}{\mathbf{E}}
\newcommand{\pp}{\mathbf{p}}

\newcommand{\LAG}[3]{L_{#1}^{#2}\left(#3 \right)}
\newcommand{\COE}{{\mathcal{C}}}

\begin{document}

\title{Spin  exciton  in  a  quantum dot
with spin-orbit coupling at high magnetic field}

\author {P.Lucignano$^{1,2}$} \author{B.Jouault$^3$} \author{A.Tagliacozzo$^{1,2}$}
\affiliation{$^1$ Coherentia-INFM, Unit\`a di Napoli}
\affiliation{$^2$ Dipartimento di Scienze Fisiche Universit\`a di Napoli
         "Federico II "\\
         Monte S.Angelo - via Cintia, I-80126 Napoli, Italy}
\affiliation{$^3$ GES, UMR 5650, Universit\'e Montpellier $II$\\
         34095 Montpellier Cedex 5, France}

\date{\today}
\begin{abstract}
 Coulomb interactions of few ($ N $) electrons confined in a disk
 shaped quantum dot, with a large magnetic field $B=B^*$ applied in the
 $ z - $direction (orthogonal to the dot), produce a fully spin
 polarized ground state.  We numerically study the splitting of the
 levels corresponding to the multiplet of total spin $S=N/2$ (each
 labeled by a different total angular momentum $ J_z $) in the presence of
 an electric field parallel to $ B $, coupled to $ S $ by a Rashba
 term.  We find that the first excited state is a spin exciton with a
 reversed spin at the origin. This is reminiscent of the Quantum Hall
 Ferromagnet at filling one which has the skyrmion-like state as its first
 excited state. The spin exciton level can be tuned with the electric
 field and infrared radiation can provide energy and angular momentum
 to excite it.
\end{abstract}

\maketitle

\section{Introduction}
Quantum dots are semiconductor devices in which electrons are
confined to a small area within a two dimensional electron gas by
properly biasing metal gates added to the
structure\cite{kouwe,jacak}. In an isolated quantum dot (QD) the
confining potential gives rise to quantized single particle energy
levels. However, electron-electron interaction determines the dot
properties. In dots having a diameter of $\sim 100nm$, the level
spacing and the Coulomb energy are of the order of $ \sim 1 meV $ and
most charging properties can be included within the Hartree-Fock
($H-F$) approximation, just like in atoms. Correlation effects do
not significantly alter the charging properties, but may strongly
influence the spin properties of the confined electrons. One striking
evidence of this is the fact that Hund's rule, which is typical of
atoms, is often satisfied in dots \cite{tarucha}. However, the
reduction of the energy scale by a factor of $ 10^{-3} $ with respect
to atoms, enhances the sensitivity of the electrons in the dot to an
external magnetic  field.\\ In this paper we discuss the spin
properties of an isolated vertical QD in presence of a magnetic field
$B$ in the $z-$direction, orthogonal to the dot disk (cylindrical
symmetry is assumed). We also include the spin orbit coupling ($SO$)
induced by an electric field  along $z$.

 In the cylindrical geometry, orbital effects are
dominant\cite{ishikawa}. Indeed Zeeman spin splitting does not drive
any spin polarization in these systems and can be often ignored
\cite{spiego}.
 However correlations combine orbital and spin effects
together and
 can be probed by magnetoconductance measurements in a
pillar configuration
 \cite{wiel,schmidt}. Spin properties are
quite
 relevant to conductance, in view of the possibility of spin
blockade\cite{imamura,jouault}, Kondo effect \cite{aleiner,arturo},
or Berry phase induced tuning\cite{berry}.

 Quantum numbers
labeling the dot energy levels are the number of electrons $N$, 
the total orbital  angular
momentum along $z$, $M$, the total spin $S$ and the
 $z-$component of
the spin $S_z $.  By increasing the magnetic field
 $B$, both $M$ and
$S $ increase.  Finally, at $B= B^*$, a fully spin
 polarized $(FSP)$
state is reached.  The increasing of the total spin
 $S$ was measured
in a dot with about 30 electrons, a striking evidence
 of e-e
correlations\cite{klein}. While, in the absence of
 interactions, the
density of the $FSP$ state becomes uniform over the
 dot area, which
is contracted to a minimum, in the real case, the e-e
 interaction
tends to reduce the density at the center of the disk, by
compressing the electrons at the dot edge (see Fig.\ref{fig000}
[middle panel]).  By further increasing $B$, the electron density
reaches a maximum value. The dot becomes a so called ``Maximum
Density
 Droplet'' ($MDD$) \cite{oosterkamp}.  For larger $B$
values,the $FSP$
 state is disrupted: the dot density reconstructs
i.e.  an annular local
 maximum of the density is produced at the edge
of the dot
 \cite{rokhinson} with breaking of azimuthal symmetry at
the edge 
 (de Chamon-Wen phase\cite{chamon}).
 
 Various
numerical calculations\cite{reimann1}
 have investigated these
subsequent electronic transformations
 which appear as crossing of
levels with different quantum numbers. 
 $H-F$ calculations are known
to incorrectly favor spin-polarized
 states \cite{pfannkuche}.  Spin
density functional calculations have
 been performed for dots
including a larger number of
 electrons\cite{koskinen}. The density
functional approach, with a
 good choice of the parameters of the
potentials, can 
 reach a significative agreement with the
experiments, 
 but it may introduce uncontrolled approximations.
When the electron density is
 reduced, a Wigner molecule can be
formed. Recently, this broken symmetry 
 state has been studied 
 in
the absence of an external 
 magnetic field, using Quantum Monte
Carlo simulations, with a multilevel
 blocking algorithm which is
free of the sign
 problem\cite{egger}.
 
 In this work, we use
exact diagonalization for few electrons with
 azimuthal symmetry
\cite{jouault} to discuss the spectral properties
 of the $FSP$ dot
which is stabilized by the Coulomb interaction.
 Electrons are
confined to a two-dimensional ($2-d$) disk by a $2-d$
 parabolic
potential and interact via the full Coulomb repulsion whose
 strength
is parametrized by $U = e^2 / \kappa l $. Here $l$ is the
 magnetic
length due to the parabolic confinement in presence of a
 field $B$
along $z$ and $\kappa $ is the static
 dielectric constant.  The
confinement of the electrons in the $x-y$ plane
 implies the
presence of an electric field in the $z-$direction, provided by the
band bending of the hetherostructure.  This gives rise to the so
called $SO$ Rashba term \cite{rashba}, which can be enhanced even
further in a non-linear conductance measurement, when an extra bias
voltage $V_{sd}$ is applied to the contacts of a vertical structure.

 In the presence of $ SO $, $J_z = M + S_z $ becomes the good
quantum
 number. The $FSP$ ground state (GS) has $J_z= N(N-1)/2+S$
($J_z=25/2$ for $N=5$), while the first excited state (denoted as
$SKD$ state in the following) has $J_z= N(N-1)/2+S-1$ ($J_z=23/2$
for
 $N=5$).  The charge density is rather insensitive to the $SO$
coupling, $\alpha $.  However, we show that the $SO$ interaction
couples the spin polarization to the orbital motion determining the
spin properties of the GS and the first excited states
 in a
surprising way.  Indeed , by increasing $\alpha $, the
 expectation
value of the spin density of the GS,
 which was originally oriented
in the $z-$direction, acquires a component in the dot plane, because
the minority spin density is increased and pushed from the center of
the dot outward.  Moreover the combined effect of $U$ and $\alpha$
deforms substantially the spin density of the $SKD$. A sharp
minority
 spin polarization is present close to the dot center.  The
reversal of
 the spin polarization at the origin in the $SKD$ state
w.r. to the
 $FSP$ GS leads to an extra node in the spin density.

This situation
 is reminiscent of the case of the Quantum Hall
Ferromagnet ($QHF$)
 \cite{tycko} close to filling one. In that case,
a true magnetic
 ordering is achieved, which is characterized by full
spin polarization
 in the GS and by a topologically constrained first
excited state, the
 Skyrmion ($SK$) state, with reversal of the spin at
the origin, first
 studied in the $O(3)$ non linear $\sigma$ model
$(NL\sigma M )$  in
$2-d$ dimensions
 \cite{polyakov,rajaraman}. In Section V, 
 we elaborate
 on the analogies and differences between the
$FSP$ dot and the
 $QHF$. In the $QHF$ a topological quantization of
charge occurs. By contrast the $SKD$ state has no topological
features, because the geometrical compactification procedure described 
in Subsection V.A  cannot take place. We refer to the $ SKD $
state as a ``spin exciton'' because  there is some piling up of the  charge
at  the center of the dot  w.r.to the GS, together with   the reversal
 of the spin polarization there.

 The paper is organized as follows:

 In Sect.II we
report our results of numerical diagonalization with 
 the Lanczos
algorithm for a dot with five 
 electrons ($N=5$) close to the $FSP$
state, in the absence of 
 $SO$ coupling.
 
 In Sect.III we derive
the $SO$ matrix elements in the $2-d$ 
 harmonic oscillator basis and
discuss how the $SO$ coupling
 modifies the lowest lying energy
levels.
 
 In Sect.IV we show the spin and charge density of the
lowest 
 lying $J_z$ multiplet when the $FSP$ state is achieved.
 
In Sect.V we summarize the features of the $SK$ state in the $QHF$ 
and compare these with the ones of the $SKD$ state in a dot with
$SO$
 coupling.

A brief summary and  some conclusions  are outlined 
 in Sect. VI.\\
 There is evidence of skyrmion excitations in $ Ga \: As \: 2-d$ electron 
gas  systems close to filling one by magnetoabsorption spectroscopy
\cite{aifer}.  A sharp absorption  line could be found in
  exciting   dots   to the
$ SKD $ state, by transferral 
 of energy and angular momentum with
circularly polarized light. This  amounts to adding a spin exciton to the
dot.

\section{FSP state and dot reconstruction}
We consider  $N=5$ electrons confined in two dimensions 
 (spanned by 
the ($\rho, \varphi$) coordinates) 
by a  parabolic potential of  characteristic frequency $
\omega _d $. This is a model for an isolated  disk shaped  QD. 
  A magnetic field $B$ orthogonal to the  disk is measured in 
units  of  $\hbar \omega_c$ ($meV$), where  $ \omega _c $ is the cyclotron
 frequency.  In the absence of spin orbit coupling,
 the single particle states $\phi _{nm}$  are the 
eigenfunctions of the $2-d$ harmonic oscillator with frequency
 $ \omega _o = \sqrt { \omega _d^2 + \frac{\omega _c^2}{4} } $.  They 
are labeled by
$n,m$ (with $n \in (0,1,2,3,...)$ and $m \in (-n,-n+2,...,n-2,n)$).
 $m$ is the angular momentum in the $z$ direction:
\begin{eqnarray}
\phi_{nm}=
 \frac{e^{im\varphi}}{l \sqrt{\pi}} R_{n|m|}(t)
=\nonumber\\
 \COE_{nm} \frac{e^{im\varphi}}{l\sqrt{\pi}} 
e^{- \frac{\rho^2}{2l^2}}
\left(
\frac{\rho}{l}
\right)^{|m|}
\LAG{\frac{n-|m|}{2}}{|m|}{\frac{\rho^2}{l^2}}\:\: .\label{primo}
\end{eqnarray}
Here $\LAG{n}{\alpha}{t}$ (with  $t=\rho^2/l^2$) is the generalized 
Laguerre polynomial with $n\geq 0$ \cite{abramowitz},
$l=\sqrt{\hbar / m^*\omega_o}$ is   the  characteristic length  due to the 
the lateral geometrical
confinement in the dot  inclusive of the $B$ field effects and
  $\COE_{nm}= 
\left[ 
\frac{ \left(\frac{n-|m|}{2} \right)! }
{ \left(\frac{n+ |m|}{2} \right)!  }
\right]^{\frac{1}{2}}
$ is a normalization factor.
  
\begin{figure}
\includegraphics*[width=0.7\linewidth]{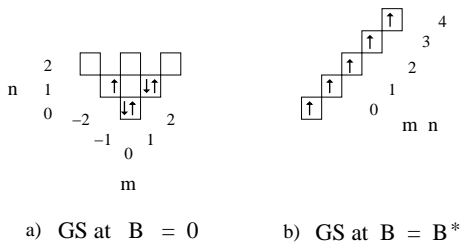}
\caption{Slater determinants  quoted in the text are depicted.
 Quantum numbers are $N=5$, $S= 1/2 $ for the state at $B=0$ $[a)]$
 and $S= 5/2$ for the state at $B=B^* $, the magnetic field value at
 which the maximum of  $S$ is achieved $[b)]$. }
\label{fig0}
\end{figure}

\begin{figure}
\includegraphics*[width=0.70\linewidth]{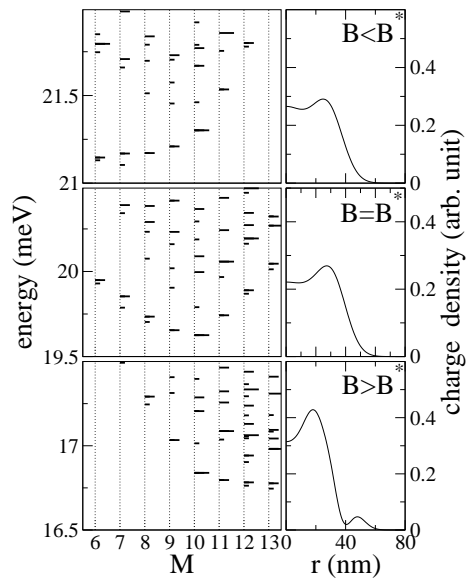}
\caption{
Energy levels without $SO$ coupling for   the dot with $N=5$ electrons
at  $U=13\:meV$ and  $\omega _d =\:meV$. Magnetic field values are: 
 (in units of $\hbar \omega _c $):  
$B=5\:meV$ [top],$B=B^*=7\:meV$ [middle], $B=11.5\:meV$ [bottom].
The total $M$ is on the $x$ axis. The levels  are drawn with short,
medium or long dashes,  depending on the total spin : $S=1/2, 3/2,5/2 $}
\label{fig000}
\end{figure}

The corresponding single particle energy levels are 
\begin{equation}
 \epsilon _{n ,m} =
(n +1) \hbar \omega _o - \frac{m}{2} \hbar \omega _c  \:\:   .
\label{elev}
\end{equation}
  In the absence of both
interaction and magnetic field, the lowest lying single particle states
 are occupied with the minimum spin. The GS
Slater determinant  is sketched 
pictorially in Fig.\ref{fig0}a,  where  energy is intended on the 
vertical  axis.
  Each box represents a single particle state labeled by
$n,m$  and arrows represent electron occupancy with spin projection
along the quantization axis. 
We have performed exact diagonalization of this system including
Coulomb interaction between the electrons. The matrix elements of the
unscreened Coulomb interaction use the  single particle
basis set, up to 28 orbitals. They can be calculated analytically and
are parametrized by the strength of the interaction $U$. 
Our calculation is limited to very small particle numbers ($ N < 7
$),because the truncation of the Hilbert space influences the results
for larger $N$.  However, our convergency checks show that the
numerical errors proliferate only at higher energies.  In particular
they affect the reliability of the level spin degeneracy. In any case,
numerical errors are quite small if is $N=5$.  In Fig.\ref{fig000}
[left panels], we show the lowest lying total energy levels at fixed
angular momentum $M$, versus $M$, for $U= 13\:meV$. Magnetic field is
$B=5\:meV$ [top], $B= B^* = 7\:meV$ [middle], $B=11.5\:meV$
[bottom]. At each $M$, the spin degeneracy is marked by dashes of
different length: short dashes for $S=1/2$ (doubly degenerate level),
medium dashes for $S=3/2$ (fourfold degeneracy) and long dashes for
$S=5/2$ (sixfold degeneracy).  On the r.h.s. of the picture the radial
charge density of the corresponding GS is plotted $vs$ distance $r$
from the dot center. Fig.\ref{fig000} ([left panels]) shows the
crossing of levels with increasing $B$. Electron-electron correlations
imply that when $M$ increases, $S$ also increases.  At $B= B^* =7\:
meV $ the spin $S$ reaches its maximum value $S=N/2$.  The largest
contribution to the GS wavefunction is given by the Slater determinant
depicted in Fig.\ref{fig0}b) corresponding to $M = \sum_0^{N-1} m = 10
$.  We concentrate on the state at $B=B^*$,the $FSP$ GS. This
corresponds to the ``maximum density droplet'' state discussed in the
literature
\cite{oosterkamp}.
Qualitatively we can say that at $B=B^*$ the dot attains its smallest
radius. As can be seen from the GS charge density, further increase of
$B$ leads to the so called reconstruction of the charge density of the
dot. For $B> B^*$, the $M$ of the GS increases further, but $S$ is no
longer at its maximum. In the bottom panel of Fig.\ref{fig000} it is
shown that at $B= 11.5 \: meV$ the GS energy is now achieved for
$M=13$ with a doublet ($S=1/2$) state.  The corresponding charge density
of the dot, as depicted on the r.h.s, is strongly modified close to
the edge\cite{rokhinson}: it displays a node followed by an extra non
zero annulus at a larger distance. In view of the fact that our
expansion of the wavefunction only includes rotationally invariant
components, the breaking of the azimuthal symmetry is impossible.  
By contrast this is found to occur in density functional calculations
and the corresponding GS is referred to as the de Chamon-Wen phase
\cite{chamon}.
The GS at $ B=B^* $ can be compared with a $FSP$ quantum Hall state of
an extended disk in the absence of lateral confinement (Quantum Hall
Ferromagnet (QHF) at filling one). Fig. \ref{fig0}b) recalls the
occupancy of the lowest Landau level (LLL) up to a maximum $m=N-1$,
except for the fact that in our case the single particle levels
corresponding to the LLL are not all degenerate in energy.  In the
language of the quantum Hall effect the unperturbed levels are:
\begin{equation}
 \epsilon _{\nu ,m} = ( 2 \nu + |m| +1 ) \hbar \omega _o - \frac{m}{2}
 \hbar \omega _c 
\label{LLL}
\end{equation}
where $ \nu = (n-|m|)/2 $.  LLL is for $\nu = 0 $ and $m \geq  0 $.  The Slater
determinant of Fig.\ref{fig0}b) has a charge density which is flat as
a function of the radius $r$, up to the disk edge, where it rapidly
falls down to zero. In our case this feature is lost because of the
presence of $U$, together with the fact that the  number of electrons
is small.  We will better discuss  the comparison of the $FSP$ GS with the $QHF$ in
Section V.
\section{Inclusion of spin-orbit}
We now add the spin orbit interaction to the $FSP$ GS at $B=B^*$.  This
can be tuned by applying an electric field $\EE$ in the $\hat z$
direction, which couples to the spin of the electrons in the dot with
a term \cite{rashba}:
\begin{equation}
H_{so}= \frac{\alpha}{\hbar}  \left( \hat {z} \times \pp \right)
\cdot \vec{\sigma}\: .
\end{equation}
Here $\vec{\sigma}$ are the Pauli matrices, $\alpha$ is the spin-orbit
coupling parameter which is
 proportional to the electric field.
$\alpha$ will be 
 measured in units of $meV\cdot \AA$.  We now
rewrite the spin-orbit coupling term in a second quantized form.
 We
denote the fermion operators associated to $\phi _{nm} $ of
Eq.(\ref{primo})
 by $c_{nm\sigma },
 c^\dagger_{nm\sigma }$ and we
get: 
\begin{widetext}
\begin{eqnarray}
\!H_{so}=
  \alpha 
\sum_{nm} \sum_{n'm'} \left \{
<n'm'| -( \drond{x} + i \drond{y}) |nm> c^\dagger_{n'm'\downarrow}
 c_{nm\uparrow}
+
<n'm'|  \drond{x} - i \drond{y} |nm> c^\dagger _{n'm'\uparrow}
 c_{nm\downarrow} \right \}\:\:  .
\end{eqnarray}
\end{widetext}
The integration over the azimuthal angle $\varphi$ can be done analytically.
  This shows
that the Hamiltonian can be rewritten in the following way:
\begin{eqnarray}
H_{so}= \frac{\alpha}{l}
\sum_{n n'} \sum_{m} \left(
B_{n'm+1,nm}  
c^\dagger _{n'm+1\downarrow} c_{nm\uparrow}  
+  \right .  \nonumber\\
\left .  A_{n'm-1,nm}
c^\dagger _{n'm-1\uparrow} c_{nm\downarrow}
\right)
\end{eqnarray}
with  $A_{n'm'nm}= $
\begin{equation}
 \delta_{m'+1,m} \: \int_0^\infty dt
R_{n'|m'|}(t) (2\sqrt{t}\drond{t}+ \frac{m}{\sqrt{t}}  ) R_{n|m|}(t)
\nonumber
\eneq
and $B_{n'm'nm}= $
\beq
 \delta_{m'-1,m} \:\int_0^\infty dt
R_{n'|m'|}(t) (2{\drond{t}}^\dagger \sqrt{t}+ \frac{m'}{\sqrt{t}} ) R_{n|m|}(t)
\nonumber
\eneq
Here $B_{nm,n'm-1} = A^*_{n'm-1,nm}$, what implies that the hamiltonian
is hermitian. It is clear that while $s_z $ and $m$  are no longer 
separately conserved,  their sum $ j_z= s_z + m$ ( with $ j_z$ half integer )
  is a good quantum number.
We will denote the single particle basis that diagonalizes the $SO$ term
by $w_{j_z}^\beta $ with $\beta = p,m $. The label $\beta $ 
takes two possible values, say $p,q $ and allows for conservation
 of the  number of degrees of freedom.   

The $SO$ interaction lifts the spin degeneracy. In 
Fig.\ref{deg2}  we show the splitting of the  multiplet  with $N=5$,
$S= 5/2$,  $M=10$  at $B=B^*=7 \:meV $ and 
$U = 13 \: meV$ {\sl vs} the strength of the $SO$
 coupling $\alpha$. The strength of $U$ is responsable not only for the fact
 that the  GS belongs to this multiplet, but also for the ordering in 
energy of the sequence: 
$J_z = 25/2,23/2,21/2,19/2,17/2,15/2 $ ( from  bottom  to top). 
At small $U$ values the sequence is $J_z =15/2,17/2,19/2, 21/2,25/2,23/2$, 
as shown in  Fig.\ref{fig3ps}. With increasing of  $U$,
 some level crossings occur. 
\begin{figure}
\includegraphics*[width=0.7\linewidth]{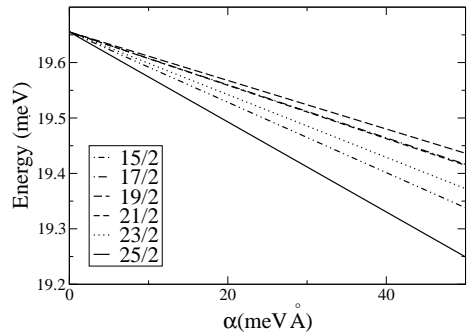}
\caption{Splitting of the lowest lying multiplet for $N=5$,$S= 5/2$ and 
$M=10$ $vs$ strength $\alpha $ of the $SO$ interaction, at 
$B=B^* =7\: meV, $ $U=13\:meV$ and $\omega _d = 5 \: meV$.  The levels are
labeled by $J_z$.}
\label{deg2}
\end{figure}
 The ordering at three different values of
$U$ is magnified in the bottom panels of  Fig.\ref{fig3ps}. 
 The case with $U = 13 \: meV $
is shown in the bottom right panel of Fig. \ref{fig3ps}. The lowest
state in energy is for $J_z=25/2$ followed by $J_z = 23/2, 21/2$
(almost degenerate with $15/2$), $15/2, 19/2, 17/2$.  At $U=13\:meV$ a
sizeable gap is formed between the $J_z=25/2$ GS and the first excited
state $J_z = 23/2$ ($SKD$).  The other states of the multiplet are
bunched together at higher energy.
\begin{figure}
\includegraphics*[width=0.7\linewidth]{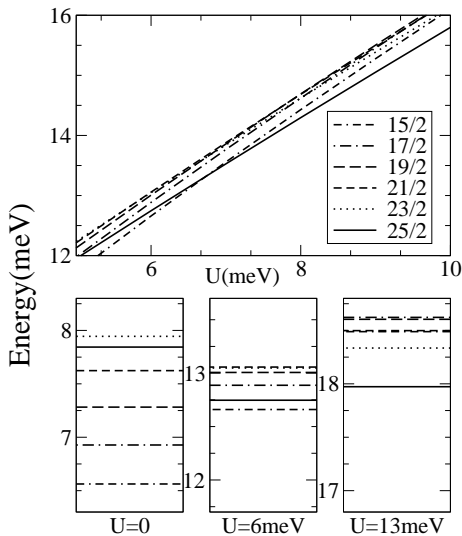}
\caption{Energy levels  with  $B=7\:meV$, $\omega _d = 5\:meV$, and 
$\alpha = 100\:meV \cdot \AA$, for different $U$ values.
 In the upper panel we
show the crossings that allow the $FSP$ polarized state to be the ground
state when $U$ is large. Ordering of the levels is magnified in the
bottom panels for three different $U$ values.}
\label{fig3ps}
\end{figure}
In this Section we focus on the $U=13 \: meV$ case and discuss the
charge density and the spin polarization density of the GS at $J_z =
25/2$. The other states of the multiplet will be analyzed in Section IV.

As it appears from Fig. \ref{fig4ps} [top panel], the charge density 
of the GS is only  mildly changed  when we  increase the $SO$ coupling.
\begin{figure}
\includegraphics*[width=0.7\linewidth]{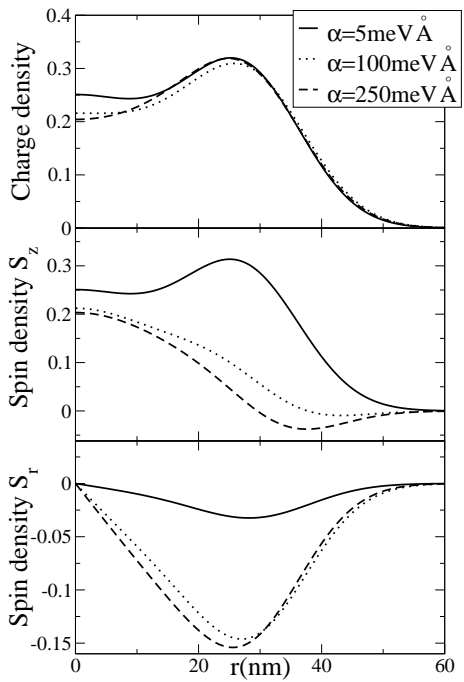}
\caption{Charge density, azimuthal spin density $S_z$, in plane spin density
$S_r$, in the radial direction, in the GS ($J= 25/2$) at various $SO$
couplings: $\alpha = 5,100,250\:meV\cdot \AA$. Here $ B= 7\:meV$, $U = 13\:meV$ and
$\omega_d = 5meV$}.
\label{fig4ps}
\end{figure}
By contrast, the spin density is quite sensitive to the addition
of $SO$, up to saturation. Now the $z-$component of the total spin
is
 no longer a good quantum number and some admixture with down spin
electrons  appears.  Indeed the role of the Rashba term is to
rotate the average electron spin. In particular, down spin electrons
are pushed away from the center of the dot, giving rise to the spin
density components $S_z(\vec r)$ (orthogonal to the dot plane), and
$S_r(\vec r)$ (in the plane of the dot), which are plotted in
Fig.\ref{fig4ps} [middle and bottom panels, respectively]. It is
remarkable that the spin density $S_z(\vec r) $ changes sign at the
edge of the dot for large $SO$ coupling.  This is confirmed by a plot of 
 the occupation numbers $n_{nm\sigma }=
 \langle
GS |c^\dagger _{nm\sigma }c_{nm\sigma }| GS\rangle $ with
$n=m$. They are shown in Fig.\ref{figN4e} for both $N = 4$ and $N =
5$ for comparison. Of course, the change of $N$ would also imply an
effective change of the confinement potential $\omega_d$ (what we do
not do). However, all what we want to show here  is that our 
findings depend on  the strength of $B$ only, and not on  the number of
electrons being even or odd.  A similar feature occurs  in the de
Chamon-Wen phase, in the absence of $SO$: when crossing the edges,
the spins 
 tilt away from their bulk direction\cite{karlhede}.
 
The
reversal of the spins in the tail at the dot boundary is a 
peculiarity of the Rashba interaction, but  the spin/charge density
 is very small there and
does not 
 influence the dot properties. 
\begin{figure}
\includegraphics*[width=0.7\linewidth]{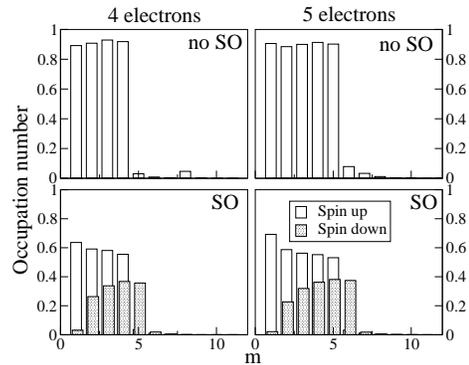}
\caption{
Occupation numbers $ n_{n=m,m,\sigma}$ in the $ GS $ with $N=4(5)$
electrons (left(right)) without $SO$ (top) and with $SO$ ($\alpha =
100\:meV \cdot \AA$) (bottom).
  Other parameter values are $B=7\: meV$, $U=13\:
meV$, $\omega_d = 5\:meV $.  White bars refers to spin up, grey bars
refer to spin down. The $FSP$ GS of the dot with $N=4(5)$ electrons
has total spin $S = 2(5/2)$ and the $z-$component of the total angular
momentum $J_z = 8(25/2) $.}
\label{figN4e}
\end{figure}
\section{Spin and charge density in the multiplet $S=5/2$, $M=10$} 
In the previous Section we have shown that at $B=B^*$ the GS with 
$N=5$ electrons belongs to the 
 $S=5/2$, $M=10$  multiplet. The $SO$ coupling lifts its degeneracy as shown in
Fig.\ref{deg2}.
The size of $U$ strongly influences the energy of each state, by
producing crossings of levels. 
 At $U = 13 \:meV$ the lowest lying states with increasing energies are
 (see Fig.\ref{fig3ps}[right bottom panel]):
\begin{description}
\item $ |GS> \equiv |N=5;\: S=5/2, J_z = 25/2 > $:  this is the fully spin 
polarized GS . 
\item $ |SKD > \equiv |N=5;\: S=3/2, J_z = 23/2 > $: the `spin exciton`.
\item $ |b> \equiv |N=5;\: S=3/2, J_z = 15/2 > $:  this is a  state 
higher in energy  w.r. to $|SKD> $.
\end{description}
This ordering of energy levels is again a consequence of Hund's
rule: lowest energy is for $J_z = L_z+|S_z|$, higher
energy is for $ J_z = L_z-|S_z|$. Besides affecting the energy of the states,
the effect of $U$  is to enhance the transfer of weight from the majority
 (``up'') to the   minority (``down'')  spin population.
 This is shown in Fig.\ref{plotN5J7.5},  where the occupation numbers 
 $ n_{n=m,m,\sigma}$ are reported for the states 
$ |GS> $, $ |SKD > $ and  $ |b> $ for $U=0 $ [left panels] and $U= 13 \:meV$
[right panels], respectively.  A striking feature characterizes the
spin densities  of these states (see  Fig.\ref{plotN5J7.5},
\ref{fig6ps}):
the dominant spin density is reversed in the $| b> $ state,
 w.r.to
the $|GS> $. The state $|SKD > $, which is the first excited state, 
interpolates between the two.  Spin occupancy is not significantly
modified for
 larger $r$.  While at $U=0$ the flipping of the spin at
the origin w.r. to
 the GS is full, in the interacting case some
up-spin is left at the
 center. This allows for a smoother radial
dependence of the spin and 
 charge density expectation
values. Eventually, this is the reason why
 this state turns out to be
the lowest excited state in the $FSP$ system.
\begin{figure}
\includegraphics*[width=0.70\linewidth]{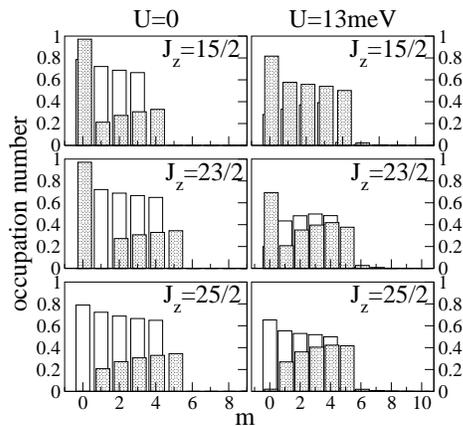}
\caption{The occupation numbers $n_{n=m,m,\sigma }$
in the state at $J_z = 15/2,23/2,25/2$ for small $U$ [$left$], and
large $U$ [$right$]. White bars refers to spin up, grey bars refers to
spin down. We stress that at $U=0$ the ordering of the levels, 
corresponding to the three panels on the left is changed  w.r. to the 
ones on the right ($U=13\:meV$).  (see Fig. \ref{fig3ps} [bottom panels]).}
\label{plotN5J7.5}
\end{figure}

 In Fig.\ref{fig6ps} we show the charge and spin densities of the
complete multiplet at $\alpha = 100\:meV\cdot \AA $, $U = 13\:meV$ and
$B = B^*
 $. The situation is quite peculiar: by looking at $ <S_z>$
[middle panel],
 we see that 
 the GS has an up spin density
everywhere in the dot, except for a 
 tiny little reversed tail at the
boundary.  By contrast, 
 the state $J_z = 15/2$ has a down spin
density at any $r$. Intermediate between the two, 
 the $SKD$ state
displays a
 reversed spin at the center of the dot but the spin
polarization
 changes into up when approaching the edge, to restore
the spin density 
 of the $25/2$ state.  There is a node in the
middle.  The other states
 ($17/2$, $19/2$ and $21/2$) are rather
featureless and they do not share
 these features.  The trend is
confirmed by looking at the
 projection of the spin density in the
plane of the dot $ S_r =
\hat{r}\cdot \vec{S} $ (see Fig.\ref{fig6ps}[bottom panel]).  This is the
complementary information w.r. to $S_z(r)$. When $S_r(r)$ in strongly
non zero, then $S_z(r) $ is heavily reduced.

\begin{figure}
\includegraphics*[width=0.7\linewidth]{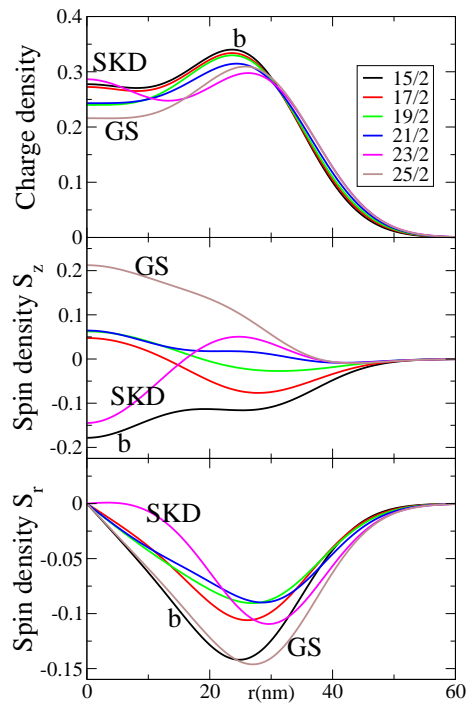}
\caption{(\sl{color on line}) Charge density, azimuthal spin density $S_z$, in plane spin
density $S_r$, in the radial direction, at various $J_z$.  From bottom
to top: $J_z = 25/2 (GS) ,23/2 (SKD) ,21/2,19/2,17/2,15/2$. Other
parameters are $\alpha = 100\:meV \cdot \AA $, $U = 13\:meV$ and $ \omega_d =
5\:meV$}.
\label{fig6ps}
\end{figure}

An analogous interpolation occurs for the charge density.
 There is a piling up of the charge at the 
origin (see Fig.\ref{fig6ps}[top panel]), 
corresponding to a locally  dominant down spin density.
The $|SKD>$ state  is a collective excitation of the QD, which we  call
 a ``spin exciton''. In the next Section 
we show that the spin exciton recalls  the first excited state of a
$QHF$ with some important differences, though.
\section{Comparison between the dot  and a QHF  disk}
The case of the dot in the $FSP$ state can be compared with that of a
disk shaped Quantum Hall Ferromagnet  at filling one.
The comparison  is in order, because the physics of the dot turns into 
that of a quantum Hall disk by increasing the magnetic field,
as long as the ratio $ \omega _d / \omega _c  \to  0$. Of course, 
while the infinite quantum Hall system is marked by  a phase transition 
to the spin polarized state, the dot, being a  system  with a finite number 
of particles, 
  undergoes  a crossover 
to the $FSP$ state  which is not a broken symmetry state.  This is confirmed
by the presence of the tiny  minority spin tail at the edge of the dot.

  In the  Subsection V.A, we recall some properties of the 
Hartree Fock description of the GS and first excited state of the QHF, which
applies to filling close  to (but less than) one. 

Similarly,some analytical approximations leading to  a simplified 
 $H-F $-like approach for the dot with $SO$ coupling
will be discussed in Subsection V.B to highlight the analogies between the
 two systems.
\subsection{Quantum Hall Ferromagnet}
In describing the QH state on a disk it is customary  to label one particle 
states with $ \nu = (n-|m|)/2 $  and $m, \sigma $, corresponding to 
the eigenvalues  $\epsilon _{\nu,m,\sigma}$ given in Eq.(\ref{LLL}).
The LLL  includes the wavefunctions $\phi _{nm} $ given by Eq.(\ref{primo}) 
 for $\nu = 0 $ and $m \geq  0 $. In this case all Laguerre polynomials
  $
\LAG{0}{|m|}{t} =1 $.  If there is no confinement potential
 ($\omega _d = 0 $), all $\epsilon _{0m}$ are degenerate.  We rename the 
LLL wavefunctions  $f_{0m} \chi_{\sigma} $
(here $\chi_\sigma $ denotes the spin $1/2 $ wavefunction)  and associate 
 the single particle fermion operators  $\hat a_{\nu=0 m \sigma }$  to them. 
In the $QHF$  at filling one, 
 the LLL subband with, say, spin up, is fully occupied:
the GS is a fully polarized spin state:
\beq
\biggl | QHF,0\biggr \rangle = \prod _{0\leq m\leq N-1}
 \hat{a}^\dagger _{0 m \uparrow }
\biggl | vac \biggr \rangle
\label{gro}
\eneq
Here $ | vac \rangle $ is the vacuum state.
 The lowest lying branch of excitations of the QHF
are spin waves.  These involve electrons in the down spin LLL subband
and holes in the up spin LLL subband.
  
It was pointed out long ago \cite{halperin} that,
 if the filling is slightly less than
one, the first excited state can be  a very special collective excitation
with $S<N/2 $ and an extra node in the spin density. The  spin
polarization is reversed  at the center, but gradually  heals to the 
dominant spin background over  a distance of many  magnetic lengths
($SK$ state). This excitation can be
traced back to the {\sl skyrmion }, the topological excitation of the
{\sl $O(3)$  $NL\sigma M$  in $2-d$}\cite{polyakov}.
A disk of  infinite radius   in coordinate space can be compactified to a 
sphere $ S^2 $ in 
${\cal{R}}^3$  having the origin in the south pole and the point at infinity
 in the north pole.  A similar compactification can be performed in the order 
parameter configurational space.  An uniform magnetization ``up'' 
is represented  by a vector pointing to the north pole everywhere on 
$S^2 $. The skyrmion is a finite action configuration 
on $ S^2$, satisfying the  classical eq.s of motion for the magnetization 
of the
 $ NL\sigma M$, conserving $ \vec{J} = \vec{S}+\vec{M} $ and belonging to a 
non trivial homotopy class.      If the topological charge is $Q = 1$,  the 
shape of the magnetization field is $\vec{s}(\vec{r}) = \hat r $,
 where $\hat r $ is 
the normal to $S^2$  at each point. $Q$ is the flux  of $\vec{s}
(\vec{r})$ through the sphere of
 unit radius.  The spin polarization is ``down'' at the 
south pole  and turns over continuously in space, until it reaches 
``up'' at the north pole.  That is, the spin polarization is flipped at the 
origin of the disk w.r.to the GS 
and turns smoothly over  away from it in the radial direction. 

 Within
Hartree-Fock \cite{fertig}, the Slater determinant $|S, K\rangle $
that describes this state conserves total $J_z$.  To construct it, a
canonical transformation is performed on the fermion operators:
\bea 
\hat{q} _j &=&
 u_j \: \hat{a} _{0 j-\met \uparrow } + v_j \: \hat{a}_{0 j+\met
 \downarrow }\: , \hspace*{0.8cm} j\in (\met , ...\infty ) \nonumber\\
\hat{p} _ {-\met} &=&\hat a _{0 0\downarrow}\hspace{5.3cm}\nonumber\\
\hat{p} _j&=&
-  v_j \: \hat{a} _{0 j-\met \uparrow } + u_j \: \hat{a} _{0 j+\met
 \downarrow } \: ,
\hspace*{0.5cm} j\in (\met , ...\infty )\:  ,
\label{can}
\enea
Normalization requires that $ |u_j|^2 + |v_j|^2 = 1 $.  Note that the
operator $\hat{p} _ {-\met} $ still belongs to the LLL as it destroys
a particle in the $f_{\nu =0, m=0}\chi_{
\downarrow}$ state. We denote by $f_{j}^{p/q}$ the single particle orbitals 
corresponding to the operators of Eq.(\ref{can}) and we use them in 
Appendix A.

 The generic Slater determinant built by means
of these operators is:
\beq
|S, K \rangle = \prod _{j =\met}^{\infty} \left
(\hat{p}^\dagger_{j-1}\right )^{n_{j-1}^p} \left
(\hat{q}^\dagger_{j}\right )^{n_{j}^q} |vac \rangle
\label{sk}
\eneq
$n_{j}^{\beta}$ are the occupation numbers of the single particle
states ($n_{j}^{p}= < p_j^\dagger p_j >, n_{j}^{q}= < q_j^\dagger q_j > $),
 with $ \sum _{j\: \beta} n_{j}^\beta  = N $.  The state of
Eq.(\ref{sk}) is labeled by the total spin $S$ and by $K$.  $S_z $ is
no longer a good quantum number and is substituted by
\beq
K = S - \met \sum_{j = \met}^{N/2} ( {n_{j}^q} -{n_{j-1}^p} )
\eneq
The state of Eq.(\ref{sk}) with $ S= N/2, K=0 $ is the $FSP$ $QHF$ ground
state of Eq.(\ref{gro}), if the only non zero occupation numbers are $
n_{j}^q =1 $ for $j\in (\met,...,N/2) $ with $ u_j = 1$ for $j\in
(\met,...,N/2) $.  This state corresponds to the $FPS$ GS of 
Fig.\ref{fig0}b) for the QD case.
\\For the hard core model the HF equations can be solved
analytically\cite{fertig}.  The lowest lying skyrmion state is $|N/2,
1 \rangle $, with
\beq
|u_j | ^2 = 1 - |v_j |^2 = \frac{\xi ^2 }{\xi ^2 + (j +\met )}
\label{skv}
\eneq
leading to the spin density $\vec{s} (\vec r )$ defined in terms of
the arbitrary length scale $\xi$ ( $r^2 = x^2+y^2$)\cite{rajaraman}
(see Appendix):
\beq
s_x (\vec{r}) = \frac{2x \xi}{r^2+ \xi ^2 };\:\:\:\:\: s_y
(\vec {r}) =\pm \frac{2y \xi}{r^2+ \xi ^2 };\:\:\:\:\: s_z
(\vec{r}) = \frac{r^2 - \xi^2}{r^2+ \xi ^2 }\: .
\label{bel}
\eneq
The $\pm $ refer to the sign of the topological charge $Q = \pm 1 $.
In the real $QHF$ the length $\xi $ is governed by the relative strength 
of the Zeeman and the Coulomb energies. 
\subsection{Dot with spin-orbit coupling}
In this Subsection we give arguments supporting our claim that the
state $ SKD $ of Section IV corresponds to the state
 $|N/2, 1 \rangle
$ in the $QHF$ limit, that is in the limit of zero 
 confinement
potential and filling one.
 Indeed, the radial distribution of the
spin density of the state $SKD$
 recalls the one of Eq.(\ref{bel})
except for a very shallow tail at the
 boundary. Away from the center
the spin polarization of the $SKD$ state 
 lines up gradually with the
one of the GS as it happens for the case of 
 the Skyrmion. As in the
$SK$ case, 
 $S_z(r)$ has an extra node at $r=\xi$.  In the presence
of $SO$,
 the length scale $\xi$ is no longer arbitrary, but is fixed
by the
 strength of the $SO$ coupling.
 
 In the case of the $QHF$ on
a disk, both rotations in real space around the 
 $z-$ axis and
rotations in spin space are good symmetries, so that 
 $M$ as well as
$S_z $ are
 conserved. This implies that an allowed $SK-$like excited
state of the real system has to be obtained by projecting the state of
Eq.(\ref{sk}) onto the subspace of definite $M$ and $S_z$. This is not
necessary in the QD with $SO$ interaction, because the $SO$
hamiltonian term only conserves $J_z$ as the state $|S,K\rangle $
does. In the following we show that a simplified $H-F$-like approach
for the dot 
 case with $SO$ coupling shows features similar to the
ones 
 described by Eq.s (\ref{can}), (\ref{skv}) and Eq.s (\ref{sk}),
(\ref{bel}). Let us first discuss $SO$ coupling in the dot at $ U = 0
$. The vector space required to diagonalize the $SO$ coupling and to
obtain 
 the eigenfunctions $w_{j_z}^\beta $ exceeds the LLL space
enormously  
 (in practice we always use the basis of Eq.(\ref{primo})
and never calculate 
 the $w_{j_z}^\beta $'s explicitly).
 As a
simple analytical approximation, we can restrict 
 ourselves to the
LLL for sake of simplicity. We have checked numerically that 
 this
approximation is largely satisfactory away from the level crossings.
In this case, diagonalization 
 of the $SO$ interaction factorizes the
problem into a collection of 
 $ 2\times 2$ matrices.
 What the $SO$
does is to mix single particle states with different $m$ and opposite
spins in the way that the transformation of Eq.(\ref{can})
shows. Indeed, $j_z$ ($j_z \equiv j $ in the following) is
conserved. Within the LLL, two $(m,
\sigma )$ values contribute to each half integer $j$: $ (m ,\uparrow
)$ and $(m+1 , \downarrow ) $.  The unperturbed energy levels involved,
$ \epsilon _{0m}$ and $ \epsilon _{0m+1}$, are  given by
Eq.(\ref{LLL}). Let the offdiagonal matrix element including the 
$SO$ coupling be $\alpha $. Then
the eigenvalues are:
\beq
\lambda _{j}^{p/q} = \met (\epsilon _{0 m} +\epsilon _{0 m+1} ) \pm \sqrt {
\frac{\delta ^2}{4} + \alpha ^2 }\:,
\eneq
where $\delta=\epsilon_{0m+1}-\epsilon_{0m}=
\omega _o -\omega _c /2$.
The  diagonalization implies a rotation in the 2-vector space 
 $ \{ f_{0m} \chi_\uparrow , f_{0 m+1} \chi_\downarrow \} $of an angle
$\gamma $ given by $
\tan 2 \gamma  =  -2\alpha /\delta $.
The  single particle states obtained in this way coincide with $f_j^{p/q}$
defined after Eq.(\ref{can}).
The mixing of the two states $ (m ,\uparrow )$ and $(m+1 , \downarrow
) $ is $j-$independent, within our approximations, because $\delta$
is.  This 
implies that the rotation angle $\gamma$ keeps roughly constant in
the radial direction, because average radial distribution of an
electron of angular momentum $j$ is $\sim l \sqrt{j+1}$. 
\begin{figure}
\includegraphics*[width=0.87\linewidth]{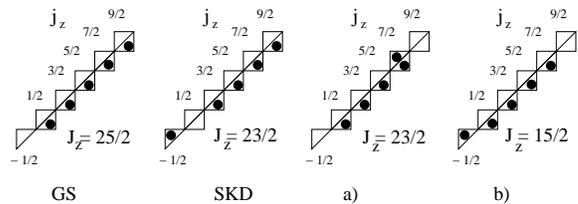}
\caption{Slater determinants  quoted in the text with the same
labels. Quantum numbers are $N=5$, $S= 5/2 $ and $J_z$. Upper$/$ lower
triangle refer to single particle states labeled by $j_z$ and
$p/q$. Other possible quantum numbers do not appear. The  dots
mark occupied states. Configuration labeled as $ a) $ is
involved in a state belonging to a much higher energy.}
\label{fig1}
\end{figure}
We can now construct the Slater determinants representing the states
lower lying in energy. The states corresponding to the ones obtained
numerically in Section IV are depicted in Fig.(\ref{fig1}). 
 In
analogy to Fig.(\ref{fig0}) we use boxes to allocate electrons. 
 Each
box is cut into a lower and an upper triangle w.r.to the diagonal, 
corresponding to the $q$ and the $p$ state of a given $j_z$,
respectively.
 A dot marks which of the orbitals is occupied.  
 We
have analyzed the Slater determinants which contribute mostly to
 the
states obtained at the end of the Lanczos procedure, giving the
average 
 occupation numbers of Fig.(\ref{plotN5J7.5}).
 Their largest
components indeed contain the 
 determinants shown in
Fig.(\ref{fig1}).

 There is close similarity with the results of
Subsection V.A.  However a relevant 
 difference can be immediately
recognized. While the skyrmion shows a very 
 smooth tilting of the
spin orientation with increasing distance from the 
 center of the
disk (see Eq.(\ref{skv})), the rotation angle $\gamma $ 
 for the dot
is uniform in the radial direction.  This feature is partly 
compensated by the addition of the Coulomb repulsion. Indeed, $U\neq 0
$
 predominantly affects the occupations close to the center of the
dot disk,
 while its influence fades out away at larger distances.
This fact introduces a radial variation of the tilting of the spin 
polarization.
 According to 
 Eq.(\ref{bel}), the skyrmion has a 
linear variation with radial distance of $ S_r $, close to the
origin.
 By contrast, 
 our numerical results reported in
Fig.(\ref{fig4ps}
 [bottom panel ]) show a quadratic increase at small
$r$'s.
 
 The role of $U$ is quite substantial, by locating the
energy of the $SKD $ state intermediate between those of the GS and 
of the $b$ state. 
 Needless to say, another
 relevant and obvious
difference between $SKD$ and $SK$ is the absence
 of any conserved
topological charge in the dot.  In
 the $QHF$ the conservation of $Q$
is implemented on symmetry grounds, by
 mapping the QH disk onto a
sphere. This mapping cannot be extended to the dot, because, as seen
from Fig.(\ref{fig6ps}),
 the direction of the magnetization at the
boundary is not unique.  Magnetization is not defined at $ \vec{r}
\to \infty$: the point at infinity is  a singular point in the magnetization
configurational space.
\section{Conclusions}
In a disk shaped quantum dot with few electrons, interactions drive
the
 system to a fully spin polarized state with $ S = N/2 $, in
 the
presence of a magnetic field $ B= B^* $ orthogonal to the dot.
 At $
B> B^* $, the total spin is again drastically reduced and the
 charge
density reconstructs at the disk boundary. We have reported
on exact
 diagonalization results of a QD with $ N = 5 $ electrons
and studied the effect of $SO$ coupling possibly due to an external 
electric field orthogonal to the dot disk.  There are analogies
between the dot state at $ B = B^* $ and the Quantum
 Hall
Ferromagnet at filling one. We require a sizeable interaction
strength $U$ to stabilize the $FSP$ GS.  When the $SO$ coupling is
increased, level crossings occur in the splitted $S=5/2$ multiplet,
until the state with maximum $ J_z = M+ S_z = 25/2$ becomes the
GS. The first excited state ($SKD$ state) has $J_z=23/2$. When
compared to the GS, the $SKD $ state has some  charge tranferred to
the dot center and a very peculiar spin texture.  Indeed, the
$z-$component of the spin density at the center of the dot is
opposite to the one of the GS and rotates continuously over away
from the center, by acquiring the same profile as the one of the GS
at the dot boundary.  This winding requires an extra node in the spin
density, which is absent in the other multiplet states. According to
these properties, the $ SKD $ state can be viewed as carrying one
spin exciton.  Both our numerical results of Section III-IV, and our
approximate analytical speculation of Section V show how essential 
the combined role of the $SO$ coupling and of the $e-e$ interaction is
 in stabilizing this state. Our calculation parametrizes the
interaction strength $U$,  but it ignores the screening of the $e-e$
interaction altogether. This should be reconsidered in view of the
fact that vertical QD's are separated on the top and the bottom from
the contact metals by barriers with a typical width of $70 \AA$. Even
for $N=5$ this is smaller than the inter-electron spacing
\cite{referee}. However, we believe that the exciton state is robust
when the screening is included. Indeed, the flipping of the spin is
concentrated at the center of the dot and is governed by the $e-e$
interaction at short range, which should be largely insensitive of
screening effects.  \\The $SKD$ state recalls the skyrmion excitation
which takes place in a disk shaped $QHF$ at filling one.  The
statement could be puzzling, in view of the fact that the $SO$
coupling is essential to the $SKD$ state, but it is never invoked when
discussing Quantum Hall properties.  However field theory models
($NL\sigma M$) use the conservation of $J$ to prove the
 existence of
the skyrmion state. In a real isolated QH disk $M,S_z$ would keep
finite and separately conserved. In this case only the component of
the $SK$ state that conserves given values of $M,S_z$ would be present
in the excitation spectrum. Nonetheless the difference is washed out
in the limit of an infinite disk size. This is the continuous limit
which leads to the $NL\sigma M$. In the case of the dot, the
compactification of both the coordinate space and the magnetization
space cannot be performed because the direction of the magnetization
is not defined at $\vec{r}\to \infty $. Therefore no state can be
constructed that conserves $J_z$ only, without conserving $M$ and
$S_z$ separately. The spin orbit coupling opens up this possibility.
However, no topological charge can be associated to the $SKD$ state in
the dot. \\Our calculation shows that for realistic values of the dot
confining potential ($\omega_d=5 meV$), of the Coulomb interaction
strength ($U=13meV$) and of the SO coupling $\alpha\sim 100meV\AA$
\cite{nitta}, the $FSP$ GS and the $SKD$ state
 are well spaced
levels. The other levels of the multiplet appear at higher energies
and are rather close to each other. This means that, at $ B = B^* $,
the dot opens a sizeable spin gap between the GS and the $SKD$ state,
that can be tuned with an applied gate.  This spin gap cannot be
washed out by thermal fluctuations, if the temperature is low enough
($\sim 50 mK$).  The gap can be
 probed by optically pumped NMR as in
quantum
 wells
\cite{tycko}. Spin-lattice relaxation of $^{71} Ga $ nuclear
spins
 in the dot, driven by the hyperfine coupling to the dot
electrons
 should be very much reduced, thus leading to a large
$T_1$.\\ The extremely low spin relaxation expected for this
excitation, could  allow for a coherent manipulation of the spin
exciton using terahertz radiation\cite{taka,awshalom,cole}. In
general, we believe that the system studied here can be relevant to
the
 coherent manipulation of  QD states. This is appealing in view
of
 quantum
 information processing\cite{chen,gchen,bonadeo}. 
Indeed, a spectrum like the one
 calculated in
 this work should
produce sharp optical absorption
 lines.
 Photoluminescence induced by
a pump and probe laser
 technique has been studied in disk shaped $
In\; Ga\; As $ QD's with
 evidence for Rabi oscillations
\cite{kamada}. In our case, because of the presence of $ B $, a
circularly polarized pulse of one single chirality can excite the
spin exciton discussed here.
\begin{acknowledgments}
The authors acknowledge important discussions with B. Altshuler, S. De
Franceschi, P. Onorato, D. Zumbuhl. 
\end{acknowledgments}

\appendix
\section{QHF Spin Density}
In this appendix we show that the  state $|N/2, 1 \rangle $ 
given by Eq.(\ref{sk}) with $u_j$ given by Eq.(\ref{skv}) leads to the skyrmion
spin density of Eq.(\ref{bel}).

The wavefunctions for the QH disk associated to the operators $
\hat{a} _{0 m \sigma } $ are given in Eq.(\ref{primo}). 
In the LLL ($\nu = (n-|m|)/2=0$) all Laguerre polynomials $
\LAG{0}{m}{t} = 1 $.  To construct the field operator, we associate a
spinorial wavefunction $ f^{p/q}_j (\vec{r} ) $ to the operator $
\hat{p}_j/ \hat{q}_j $ following Eq.(\ref{can}):
\beq
 f^p_j (\vec{r} )=
\begin{pmatrix}
 - v_j f_{0 j-\met } (\vec{r} ) \\ u_j f_{0 j+\met } (\vec{r} )
\:
\end{pmatrix} , \hspace*{0.8cm} j\in (\met , ...\infty )
\eneq
and analogously for $f^q_j$. We take $u_j $ and $v_j $ real.  The
field operator is:
\beq
\hat \psi (\vec{r}) = \sum _{j = \met }^{\infty}
 \left ( f^p_{j-1}(\vec r ) \hat{p}_{j-1} + f^q_{j}(\vec r )
 \hat{q}_{j} \right )\:.
\eneq
The spin density operator is $\hat { \vec{s}} (\vec r ) = \Re e \left \{
\hat \psi ^\dagger (\vec r ) \vec{\sigma } 
\hat \psi (\vec r ) \right \}  $, 
to be evaluated on the state $|N/2, 1\rangle $.  Let us consider
$s_x(\vec r)$ first.  The term including the $\hat{p}_j$ operators
does not contribute, because all the $f^p $ orbitals are unoccupied in
the state $ |N/2, 1 \rangle $, except for $j =-\met $.  On the other hand 
this term, does not appear, because $ u_{-\met } \cdot v_{-\met}
\equiv 1\cdot 0 =0$.

The contribution to $s_x(\vec r)$ given by the $\hat{q}_j $ operators
is:
\bea
 \sum_{j =\met } ^{\infty}
\begin{array}{cc}
 \left( u_j f^{ *}_{0 j-\met} (\vec{r} ) \right . & \left .  v_j f^{ *}_{0
 j+\met} (\vec{r} ) \right ) \\ & \end{array}
\left(
\begin{array}{cc} 
 0 & 1 \\ 1 & 0
\end{array}
\right)
\left(
\begin{array}{c}
 u_j f_{0 j-\met }( \vec{r} ) \\ v_j f_{0 j+\met } (\vec{r} )
\end{array}
\right)  \nn\\ 
=\sum_{j =\met } ^{\infty} 2 u_j v_j f^{ *}_{0 j-\met} (\vec r ) f _{0
j+\met} (\vec r )\hspace{0.5cm}.\hspace{1cm}
\enea
Using Eq.(\ref{skv}) we get $(\vec{r} \equiv (r,\varphi ))$:
\bea
&2 \xi& \sum_{j =\met } ^{\infty} \Re e \left \{ e^{i\varphi }
\frac{r^{j-\met} r^{j+\met}}{(j -\met )!^\met (j +\met )!^\met }
\: \frac{(j+\met )^\met }{\xi ^2 +  ( j + \met ) } e^{-r^2} \right \}\nn\\
 = &2 \xi& \; r \cos \varphi \: \sum_{j =\met } ^{\infty}
\frac{ (r^2)^{j-\met}   e^{-r^2} }{(j -\met )! }
\: \frac{1 }{\xi ^2 +  ( j + \met ) }\: .
\enea    
Because the maximum of the first factor occurs for $ j+\met \sim r^2 $
we evaluate the denominator of the second factor by substituting
$j+\met \to r^2 $, what allows us to perform the sum explicitly.
By noting that $ r\cos\varphi = x $ we obtain $ s_x ( \vec r ) $ as
given by Eq.(\ref{bel}).
A similar calculation applies for $ s_y ( \vec r ) $. In the case of $
s_z ( \vec r ) $, the extra factor is $v_j^2 -u_j^2 = [(j+\met ) -\xi
^2 ] / [(j+\met ) +\xi ^2 ]$ and $ \varphi $ disappears.  Using the
same approximations as above, we obtain the result of Eq.(\ref{bel}).


\begin{thebibliography}{40}
\bibitem{kouwe}
L.P. Kouwenhoven, D.G. Austing, S. Tarucha 
Rep. Prog.Phys. {\bf 64} (6), 701-736 (2001); L.P. Kouwenhoven and C.M. Marcus 
Phys. World{ \bf 116}, 35-39 (1998); M.A.Kastner, Ann.Phys.(N.Y.) {\bf 9 }, 885 (2000).
\bibitem{jacak}L.Jacak, P.Hawrilack, A.W\'ojs, Quantum Dots,
Springer-Verlag Berlin (1998).
\bibitem{tarucha}S. Tarucha, D. G. Austing, T. Honda, R. J. van der Hage, 
and L. P. Kouwenhoven,  Phys. Rev. Lett. {\bf 77}, 3613 (1996). 
\bibitem{ishikawa}Y.Ishikawa and H. Fukuyama, Journ. of  Phys. Soc. Japan {\bf 68}(7), 2405  (1999).
\bibitem{spiego} This
is due to the fact that the hamiltonian term coupling $B_{\perp} $ to
$M$ is $ \mu _B^* B_{\perp} M$ where $\mu_B^* = e\hbar/2 m^* c$ is 
the effective Bohr magneton ($m^*$ is the effective mass). 
On the other hand the
Zeeman spin splitting term is $g^*
\mu_B B_{\perp} S_z$, where $g^*$ is the effective gyromagnetic factor
for electrons in this geometry (very low in many semiconductor
heterostructures)  while $\mu_B $ is the Bohr magneton with the bare
mass (spin is insensitive to band effects). The situation is similar
to what happens in the QHE where $\omega_c$ includes the effective
mass and Landau level separation is increased by a factor of $\sim
20$, while Zeeman spin splitting, being reduced by a factor of $4$,
becomes negligible in comparison.(A.H.MacDonald,in NATO ASI:'Quantum
transport in semiconductor submicron structures', B.Kramer ed.(Kluwer
Berlin, 1996))
\bibitem{wiel} W.G. van der Wiel et al.,
Physica  B {\bf 256}, 173 (1998).
\bibitem{schmidt}T.Schmidt et al., Phys.Rev.B {\bf  51},5570 (1995).
\bibitem{imamura}
H.Imamura, H.Aoki, and P.A.Maksym, Phys.Rev.B {\bf 57},R4257 (1998);
D. Weinmann, W. Ha\"usler, and B. Kramer, Phys. Rev. Lett. {\bf 74}, 984
(1995).
\bibitem{jouault}B.Jouault, G.Santoro, A.Tagliacozzo, Phys.Rev.B {\bf 61}, 10242 (2000).
\bibitem{aleiner}  I.L. Aleiner, P.W. Brouwer and L.I. Glazman, Phys.
Rep. {\bf 358}, 309 (2002). 
\bibitem{arturo}D.Giuliano, B.Jouault, A.Tagliacozzo, Europhys. Lett. {\bf 58}, 401 (2002).
\bibitem{berry}D. Giuliano, P. Sodano, A. Tagliacozzo, Phys. Rev. B 67, 155317 (2003).
\bibitem{klein} O.Klein et al.,  Phys.Rev.Lett {\bf 74}, 785(1995);
O.Klein et al.,  Phys.Rev.B {\bf  53},4221(1996).
\bibitem{oosterkamp}T.H.Oosterkamp, et al., Phys.Rev.Lett. {\bf 82},
2931, (1999).
\bibitem{rokhinson}L.P.Rokhinson, L.J. Guo, S.Y. Chou and D.C. Tsui, 
Phys.Rev.Lett. {\bf 87}, 166802 (2001).
\bibitem{chamon}C.de Chamon and X.G.Wen,  Phys.Rev.B {\bf 49}, 8227 (1994);
S.M.Reimann, M.Koskinen, M.Manninen, B.R.Mottelson, Phys.Rev.Lett.{\bf
83},3270 (1999).
\bibitem{reimann1}for a review see: S.M.Reimann and M.Manninen, Rev.Mod.Phys.
{\bf 74}, 1283 (2002).
\bibitem{pfannkuche}D.Pfannkuche, V.Gudmindsson, and P.A.Maksym,
 Phys.Rev.B {\bf 47}, 2244 (1993).
\bibitem{koskinen} M.Stopa, Phys.Rev.B {\bf 54}, 13767 (1996).
\bibitem{egger}R.Egger, W.H\"ausler, C.H.Mak, and H. Grabert Phys. Rev. Lett. 
{\bf 82} 3320 1999; R.Egger, W.H\"ausler, C.H.Mak Phys. Rev. Lett. {\bf
83} 462 1999.
\bibitem{rashba}E.I.Rashba, Fiz. Tverd. Tela {\bf 2}, 1224 (1960) [Sov.Phys. -
Solid State {\bf 2}, 1109 (1960),Y.A.Bychkov, E.I.Rashba, J.Phys.{\bf C17},
 6039 (1984).
\bibitem{tycko} R.Tycko et al., Science {\bf 268}, 1460 (1995).
\bibitem{polyakov}A.A.Balavin and A.M.Polyakov, JETP Lett.{\bf 22}, 245(1975);
A.M.Polyakov, Phys.Lett. {\bf 59B}, 79 (1975).
\bibitem{rajaraman}R.Rajaraman {\it Soliton and instantons}, North Holland, Amsterdam (1982).
\bibitem{aifer}E.H.Aifer, B.B.Goldberg and D.A.Broido, Phys. Rev. Lett. {\bf 76},
680(1996)
\bibitem{abramowitz} M.Abramowitz, I.A.Stegun, Handbook of
Methematical Functions with Formulas, Graphs and Mathematical Tables,
Dover (New York 1972).
\bibitem{karlhede} A.Karlhede et al.,  Phys.Rev.Lett.{\bf 77}, 2061
(1996).
\bibitem{halperin}B.I.Halperin, Helv.Phys. Acta {\bf 56},75 (1983)
\bibitem{fertig}
H. A. Fertig et al., Phys.Rev.B {\bf 55}, 10671 (1997) ;
M.Abolfath et al.,  Phys.Rev.B {\bf 56}, 6795 (1997);
\bibitem{referee}We indebited to the referee for this remark.
\bibitem{nitta} In $In\;Ga\;As $ $\alpha$ is
about $70meV\AA$ at zero electric field. It can be increased by applying a gate voltage. (J.Nitta, T.Akazaki, H.Takayanagi and T.Enoki,
Phys. Rev. Lett, {\bf 78 },1335 (1997); D.Grundler, Phys. Rev. Lett. {\bf 84},
 6074 (1999)).
\bibitem{chen}P.Chen,C.Permarocchi and L.J.Sham, Phys. Rev. Lett.{\bf 87},
067401 (2001).
\bibitem{gchen}G.Chen et al., Science {\bf 289}, 1906 (2000).
\bibitem{bonadeo}N.H.Bonadeo et al., Science {\bf 282 }, 1473 (1998).
\bibitem{taka}T.Takagahara J. of Luminescence {\bf 87-89}, 308 (2000).
\bibitem{awshalom}D.D.Awshalom, J.M.Kikkawa, Phys.Today. {\bf 52}, 33 (1999).
\bibitem{cole}B.E.Cole  et al., Nature {\bf 410},60 (2001).
\bibitem{kamada}H.Kamada  et al., Phys. Rev.Lett.{\bf 87},246401 (2001).
\end{thebibliography}
\end{document}